\documentclass[
 reprint, superscriptaddress,
 amsmath,amssymb,
 aps,
10pt]{revtex4-1}

\usepackage{graphicx}
\usepackage{dcolumn}
\usepackage{bm}
\usepackage{hyperref}
\usepackage{subfigure}
\usepackage{color}
\definecolor{nrppurple}{RGB}{128,0,128}%{75,0,130}
\hypersetup{
	colorlinks=true,
	linkcolor=blue,
	citecolor= nrppurple,
	urlcolor=nrppurple
}

\begin{document}

\title{Resistivity saturation in an electron-doped cuprate}

\author{Nicholas R. Poniatowski}
\affiliation{Maryland Quantum Materials Center and Department of Physics, University of Maryland, College Park, Maryland 20742, USA}
\author{Tarapada Sarkar}
\affiliation{Maryland Quantum Materials Center and Department of Physics, University of Maryland, College Park, Maryland 20742, USA}
\author{Sankar Das Sarma}
\affiliation{Condensed Matter Theory Center and Joint Quantum Institute, Department of Physics, University of Maryland, College Park, Maryland 20742, USA}
\author{Richard L. Greene}
\email{rickg@umd.edu}
\affiliation{Maryland Quantum Materials Center and Department of Physics, University of Maryland, College Park, Maryland 20742, USA}

\date{\today}

\begin{abstract}
We report the observation of resistivity saturation in lightly doped ($x\sim 0.10)$ as-grown samples of the electron-doped cuprate La$_{2-x}$Ce$_x$CuO$_4$ (LCCO). The saturation occurs at resistivity values roughly consistent with the phenomenological Mott-Ioffe-Regel criterion once the low effective carrier density of these materials is included in the analysis. These results imply that, at least for light doping, the high-temperature metallic phase of these materials is not necessarily strange and may be understood as simply a low-density metal. 
\end{abstract}

\maketitle

The semi-classical Drude-Boltzmann theory of transport, a kinetic theory of  electronic states with well-defined momenta (``quasi-particles'') undergoing collisions, is spectacularly successful in describing the transport properties of conventional metals \cite{Ziman}. At low temperatures, transport is limited by impurity scattering, and at higher temperatures, phonon scattering dominates. However, as was originally pointed out by Ioffe and Regel in 1960 \cite{iofferegal}, the notion of a quasiparticle breaks down when the mean free path $\ell$, the average distance a quasiparticle travels before suffering a collision, becomes comparable to the lattice constant $a$. Several years later, Mott \cite{mott} re-interpreted this fact to propose a ``minimal metallic conductivity'' associated with the lower bound on the mean free path, $\ell \gtrsim a$. Since the resistivity increases with increasing temperature in metals, often linearly because of the equipartition properties of phonon occupancy, the resistivity (mean free path) can be large (small) at high temperatures approaching this ``minimal metallic'' limit.

This bound on the validity of semi-classical quasiparticle transport, now called the Mott-Ioffe-Regel (MIR) limit, is often stated in the contemporary literature in a slightly different way than the original bound $\ell \gtrsim a$. One instead argues that the quasiparticle concept breaks down when the mean free path becomes shorter than the wavelength of the electronic state $\lambda = 2\pi/k_F$, leading to the alternate statements of the bound $k_F \ell \gtrsim 1$ or $k_F \ell \gtrsim 2\pi$. In usual metals, $a$ and $\lambda$ happen to be almost equal fortuitously, but in low density systems $\lambda > a$, and the appropriate MIR criterion is $k_F \ell \sim 1$, which is the MIR criterion most commonly used in the literature \cite{sankarref}.  Typically, the longer of the two length scales, $\lambda$  or $a$, enters the phenomenological MIR criterion providing a lower saturation resistivity scale.

When the mean free path saturates this bound, i.e. when $k_F \ell \approx 1$, one can estimate the corresponding upper bound on the resistivity  $\rho(T)$ by using the Drude formula, 
\begin{equation} \label{mirbound}
    \rho_{\text{MIR}} = \frac{3\pi^2 \hbar}{e^2 k_F} \; ,
\end{equation}
where we have assumed a three-dimensional spherical Fermi surface so that $k_F = (3\pi^2 \, n)^{1/3}$, with $n$ the effective carrier density of the system. For a typical metal with $n \sim 10^{23} \; \text{cm}^{-3}$, this upper limit is $\rho_{\text{MIR}} \approx 150 \; \mu\Omega \,\text{cm}$ and for resistivities approaching this limit, quasiparticles become ill-defined as the mean free path becomes shorter than the Fermi wave length. 

For most ``good'' metals, e.g. copper where $\rho (300 \; \text{K}) \sim 1 \; \mu \Omega \,\text{cm}$, this limit is irrelevant as the temperatures needed to reach $\rho_{\text{MIR}}$ far exceed the material's melting point. The MIR limit is never reached in any normal metal where the resistivity increases linearly, but remains more than an order of magnitude below $\rho_{\text{MIR}}$. There are, however, highly resistive materials, such as alloys \cite{mooij} and A15 compounds \cite{fiskwebb}, which manifest ``resistivity saturation,'' wherein the high-temperature $\rho(T)$ plateaus at a constant value close to the predicted MIR limit $\rho_{\text{MIR}} \approx 150 \; \mu \Omega \, \text{cm}$ \cite{saturationrmp,husseyreview,satrev3}. Subsequently, high-temperature resistivity saturation has also been observed in several correlated systems, including heavy fermion compounds \cite{devisser, palstra, smithreview} and, recently, iron-pnictide superconductors \cite{bach}. Although the precise mechanism underlying resistivity saturation has not been decisively established, it is generally accepted that normal Fermi liquid type metals should not manifest a resistivity larger than $\rho_{\text{MIR}}$ even if the resistivity below this limit increases linearly with temperature as happens in most metals due to phonon scattering.  The qualitative reasoning is that the MIR limit implies sufficiently strong scattering to invalidate the existence of a Fermi surface and the associated Landau Fermi liquid picture. 

In marked contrast, resistivity measurements of the cuprate high-temperature superconductors often find non-saturating resistivities at high temperatures, and attaining values as high as $1 \; \text{m}\Omega \, \text{cm}$ or above. Specifically, a maximum value of $\rho (600 \,\text{K}) \approx 0.6 \; \text{m} \Omega \; \text{cm}$ was found in Bi$_{2+x}$Sr$_{2-y}$CuO$_{6\pm\delta}$ (Bi2201) \cite{martinbsco}; $\rho (1000 \, \text{K}) \approx 0.9 \; \text{m}\Omega \, \text{cm}$ in La$_{2-x}$Sr$_x$ CuO$_4$ (LSCO) with $x = 0.15$ \cite{lscotakagi}; and $\rho (900 \; \text{K}) \approx 1.1 \; \text{m}\Omega \, \text{cm}$ in YBa$_2$Cu$_4$O$_8$ (YBCO) \cite{bucherybco}. Since these values are all greatly in excess of $\rho_{\text{MIR}} \sim 150 \;\mu\Omega \, \text{cm}$ predicted for usual metals, these results are often taken as evidence that the cuprates violate the MIR limit, and consequently the high-temperature highly-resistive metallic phase of the cuprates is a ``bad metal'' \cite{badmetals} that does not admit a quasiparticle description. The tacit assumption here is that cuprates in their normal phase remain metallic even when the transport mean free path might be, as inferred from the  very high resistivity, shorter than the putative Fermi wavelength.

Moreover, the resistivity of these materials  exhibits a much-discussed linear temperature dependence, $\rho \propto T$ with a single slope that is unchanged from the superconducting critical temperature $T_c$ to the highest measured temperatures, as beautifully demonstrated in Bi2201 \cite{martinbsco}, as well as LSCO and YBCO \cite{lscotakagi, bucherybco}. Of course, most normal metals also manifest a well-understood phonon scattering induced linear-in-$T$ resistivity for $T > 50$ K, and therefore, it is unclear that a such a linearity itself is necessarily a strange property. Two arguments are often made based on the continuity of $\rho(T)$ over such a wide temperature range: first, that the linear-in-$T$ resistivity cannot be due to electron-phonon scattering (as occurring in conventional metals \cite{Ziman}) since it sometimes extends down to low temperatures; and second, that the lack of any discernible change in the behavior of $\rho(T)$ across the MIR limit suggests that the ``non-quasiparticle'' nature of the high-temperature phase could persist down to temperatures of order $T_c$, with potential implications for the nature of the metallic ground state and superconducting pairing in these materials \cite{badmetals}. The first of these arguments is addressed at length by one of the authors in Ref. \onlinecite{sankarlineart}, and in this work we remain agnostic about the underlying scattering mechanism in the system, while the second establishes the importance of the question of whether or not the cuprates violate the MIR limit. In the current work, we focus on this important issue of bad metallic behavior with respect to the MIR limit in cuprates.

Having briefly reviewed the literature of this field and its typical interpretation, we now make the somewhat controversial claim that the experiments described above, with the measured resistivities as high as $1 \; \text{m}\Omega \, \text{cm}$, do not constitute  \textit{prima facie} evidence that the cuprates violate the MIR limit. This is based on the simple observation that $\rho_{\text{MIR}}$ is strongly dependent on the effective carrier density, as seen in Eq. \ref{mirbound} which explicitly involves $k_F$, in conjunction with the fact that cuprates are low carrier density systems (i.e. they have small Fermi surfaces) where $n \sim 10^{21} \; \text{cm}^{-3}$. Based on these considerations, one finds that the typically quoted value of $\rho_{\text{MIR}} \approx 150 \; \mu \Omega \, \text{cm}$ for normal metals (with $n \sim 10^{23}\; \text{cm}^{-3}$) is inappropriate for the cuprates; rather, a rough estimate using Eq. (\ref{mirbound}) yields $\rho_{\text{MIR}} \sim 1-10 \; \text{m}\Omega \, \text{cm}$. In fact, measurements of lightly doped LSCO with $x = 0.04 - 0.08$ exhibit tendencies toward saturation, with $\rho (1000 \, \text{K}) \approx 2-8 \; \text{m}\Omega \, \text{cm}$, but because these values greatly exceed 150 $\mu \Omega \, \text{cm}$ this near-saturation was attributed to an unknown mechanism beyond the MIR limit \cite{lscotakagi, lscosat2}.

The largest non-saturating resistivites measured in Refs. \onlinecite{martinbsco, lscotakagi, bucherybco} only reach the lower limit of this window, and hence are not ``smoking gun'' evidence of MIR violation. Thus, the question of whether or not the cuprates violate the MIR limit is a nontrivial question which, in our view, has yet to be definitively answered.

In this work, we aim to access MIR physics by capitalizing on the fact that it is $\rho_{\text{MIR}}$ which is fundamental, not the temperature at which saturation occurs. In particular, by studying samples with large resistivities one may forego the elevated temperatures which inevitably lead to sample degradation through oxygen loss and other materials science problems. To this end, we study as-grown (un-annealed) samples of the electron-doped cuprate La$_{2-x}$Ce$_x$CuO$_4$ (LCCO) at the low doping $x = 0.08$. This enables us to approach very high ($> 1 \; \text{m}\Omega \, \text{cm}$) resistivities at easily experimentally accessible temperatures ($\sim$ 400 K) without any sample degradation complications.

The undoped parent compound, La$_2$CuO$_4$ is an antiferromagnetic correlated insulator  which becomes superconducting with Cerium doping at $x = 0.07$ \cite{musr, smrev-updated}. The samples studied in this work are thin films of LCCO ($\approx 150 \; \text{nm}$ thick) epitaxially grown on SrTiO$_3$ substrates using the pulsed laser deposition technique (see Ref. \onlinecite{taraprb} for details) and cooled in oxygen ($\sim 1 $ torr) after growth. 

By studying un-annealed (non-superconducting) films, we can achieve room-temperature resistivities $\sim 1\; \text{m}\Omega \, \text{cm}$. The $ab$-plane resistivity was measured as a function of temperature from 50 K to 400 K for a dozen as-grown samples using a Quantum Design Physical Properties Measurement System. Repeated measurements show that sample degradation is not a major concern and the measured resistivity is reproducible below 400 K, the temperature regime of interest for this study. 

\begin{figure}
    \centering
    \includegraphics[width=.45\textwidth]{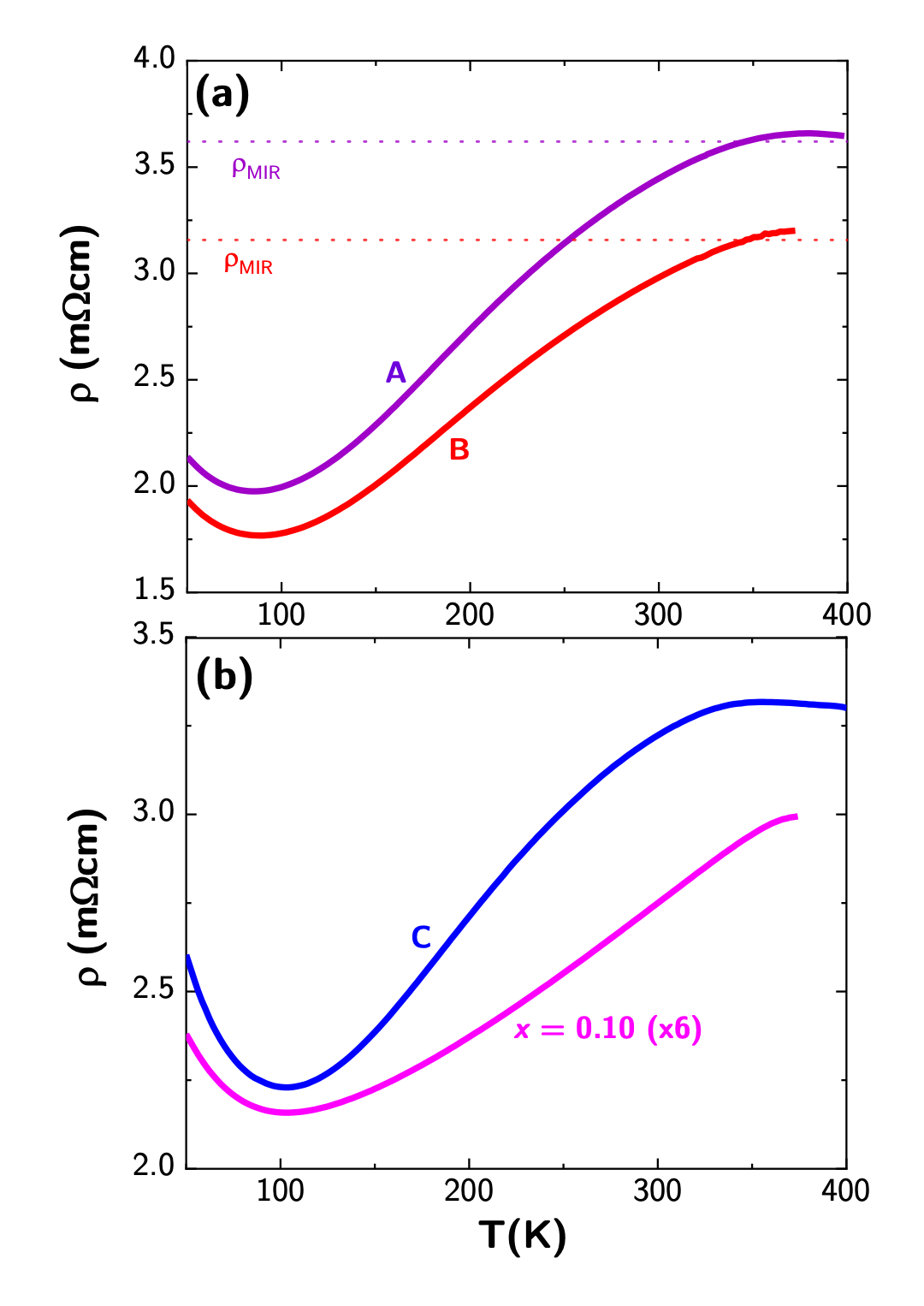}
    \caption{(a) $ab$-plane resistivity vs. temperature for two as-grown $x = 0.08$ LCCO samples. The dotted lines are estimates of the MIR limit calculated from Eq. (\ref{mirbound}) using $k_F = (3\pi n)^{1/3}$ where $n$ is the carrier density of each sample, as determined from Hall measurements at 300 K. The values of $n$ and $\rho_{\text{MIR}}$ are given in Table \ref{params}; (b) The resistivity of another as-grown $x = 0.08$ sample (blue) alongside the resistivity of an as-grown $x = 0.10$ sample (magenta), which is also seen to saturate at the highest measured temperature. Note the $x = 0.10$ curve is magnified by a factor of 6 for visibility.} 
    \label{greatsamples}
\end{figure}

Representative data for two samples are shown in Figure \ref{greatsamples}. The notable feature of these data, and the main finding of this work, is that the resistivity does indeed saturate at $\approx 3 \; \text{m}\Omega \, \text{cm}$. In fact, saturation (or near saturation) is observed for \textit{all} measured samples. We note that the value $\rho_{\text{sat}}$ at which the saturation occurs is substantially larger than the (non-saturating) resistivities reported at the highest-measured temperatures in hole-doped cuprates \cite{martinbsco, lscotakagi, bucherybco} and of similar magnitude to those found in prior studies of other electron-doped cuprates \cite{bach}. We believe that this saturation regime is not reachable in hole-doped cuprates just as the MIR limit is unreachable in normal metals.

The low-temperature upturn in the resistivity in our results is a well-known feature of the transport phenomenology of underdoped cuprates (both hole- and electron-doped) \cite{originalupturn, upturn2, pccoupturn}, and of little consequence to this work, which is focused on the high-temperature MIR regime. 

To make quantitative comparisons to the MIR criterion, we performed room-temperature Hall effect measurements to estimate the carrier density of each sample from $R_H = 1/ne$. Below 100 K, the Hall coefficient is strongly temperature-dependent, but above 200 K has only a weak temperature dependence, such that the carrier density measured at 300 K is very close to its value at 400 K \cite{taraprb}. The measured Hall values of $n$ for each sample are listed in Table \ref{params}.

Given the measured effective carrier density, we can estimate the MIR limit using the Drude formula of Eq. (\ref{mirbound}) assuming an effective spherical Fermi surface, which is a highly simplified approximation, but should suffice in view of the phenomenological nature of the MIR criterion (and the absence of a microscopic transport theory for cuprates).  Although our main experimental finding is the observation of resistivity saturation, we provide a comparison of the experimental saturated resistivity with the MIR limit using Eq. (\ref{mirbound}), obtaining a surprisingly good agreement.

The values of $\rho_{\text{MIR}}$ obtained using Eq. (\ref{mirbound}) are indicated by the dotted horizontal lines in Fig. \ref{greatsamples}(a), which agree remarkably well with the measured values of $\rho_{\text{sat}}$. In fact, for all samples measured, $\rho_{\text{sat}}$ agrees qualitatively with the calculated $\rho_{\text{MIR}}$. We note in passing that the two samples shown in Fig. \ref{greatsamples}(a) are those in which the agreement is the best, and are also the samples with the largest ratios of $\rho_{\text{sat}}/\rho_{\text{min}}$, where $\rho_{\text{min}}$ is the  minimal value of the resistivity, typically attained around $\sim$ 80 K. 

In Fig. \ref{greatsamples}(b), in addition to another $x = 0.08$ sample exhibiting resistivity saturation, we plot the resistivity of an as-grown $x = 0.10$ sample.  This sample has a substantially smaller resistivity ($\approx 0.5\; \text{m}\Omega \, \text{cm}$) than the $x = 0.08$ samples, but is nonetheless seen to begin saturating at $\sim 375$ K. We also note that the slight downturn in the resistivity above $\rho_{\text{sat}}$, which is most visible in sample C, is common to many systems which exhibit resistivity saturation \cite{saturationrmp, devisser,palstra}, thus showing that the MIR phenomenology in our cuprate system is generic.

\begin{figure}
    \centering
    \includegraphics[width=.45\textwidth]{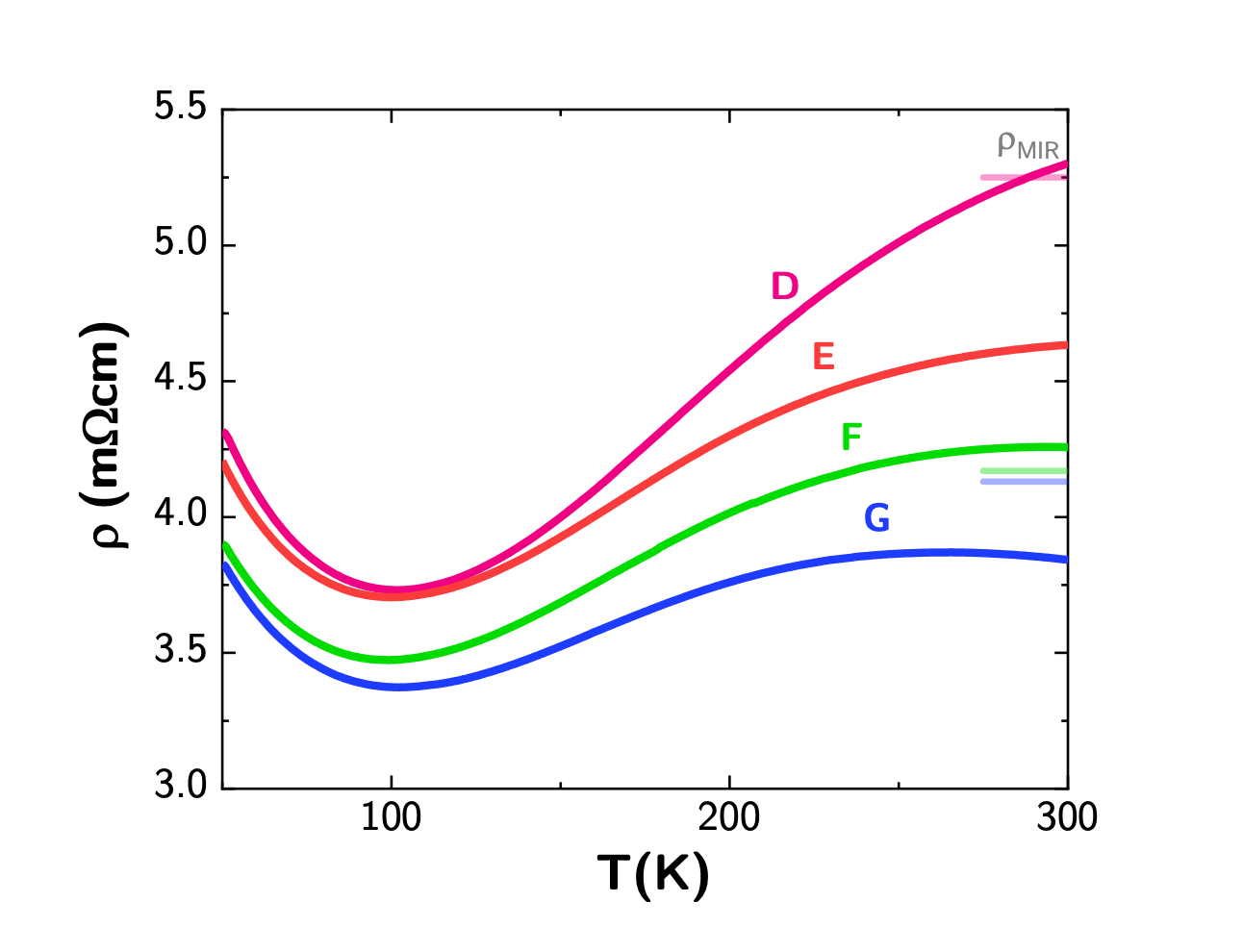}
    \caption{Resistivity traces of four as-grown $x = 0.08$ samples, all of which exhibit resistivity saturation at room temperature. The carrier densities and estimates of $\rho_{\text{MIR}}$ for each sample are given in Table \ref{params}. The transparent marks on the right of the figure indicate $\rho_{\text{MIR}}$ for each sample for which the carrier density was measured.}
    \label{othersamples}
\end{figure}

\begin{table}
\begin{center}
\begin{tabular}{ |c|c|c| } 
 \hline
 Sample & $n \; (\text{cm}^{-3})$ & $\rho_{\text{MIR}} \; (\text{m}\Omega \, \text{cm})$ \\ 
 \hline
 A & $1.28 \times 10^{21}$ & 3.62 \\ 
 B & $1.93 \times 10^{21}$ & 3.16 \\ 
 C & $2.27 \times 10^{21}$ & 2.99 \\
 D & $4.17 \times 10^{20}$ & 5.25 \\
 E & -- & -- \\
 F & $8.64 \times 10^{20}$ & 4.13 \\
 G & $8.34 \times 10^{20}$ & 4.17 \\
 \hline
\end{tabular}
\end{center}
\caption{Values of the carrier concentration for each sample shown in Figs. \ref{greatsamples} and \ref{othersamples}, as determined by Hall measurements at 300 K, as well as the corresponding estimates of the Mott-Ioffe-Regel (MIR) bounds for each, calculated using Eq. (\ref{mirbound}) as described in the main text. The carrier density of sample E was not measured. }
\label{params}
\end{table}

Further, by preparing samples with even lower carrier densities ($\sim 10^{20} \; \text{cm}^{-3}$), we can attain resistivity saturation at room temperature, as shown in Fig. \ref{othersamples} for four different as-grown $x = 0.08$ samples with slightly different densities. The values of $n$ and $\rho_{\text{MIR}}$ for each are tabulated in Table \ref{params}. It is clear that we observe resistivity saturation, and the experimental saturated resistivity is consistent with the MIR limit as estimated within a simple Dude formula.

Our finding of resistivity saturation is in apparent disagreement with much of the literature on hole-doped cuprates, where it is often claimed that the non-superconducting phase is a bad metal, based mainly on the lack of any observed resistivity saturation.  By contrast, our results presented in Figs. \ref{greatsamples} and \ref{othersamples} in the current work clearly establish the existence of the resistivity saturation phenomenon in lightly-electron-doped cuprates at a high value of $\rho_{\text{MIR}}$, as consistent with the small Fermi surface (i.e. small $k_F$) in these materials.  We believe that the possibility that hole-doped cuprates might manifest resistivity saturation if the resistivity can be increased to several $\text{m}\Omega \, \text{cm}$ cannot be ruled out; the problem is the degradation of the system at high temperatures (so that the resistivity is high) due to loss of oxygen.  The important point is that cuprates have small Fermi surfaces (i.e. low carrier densities), leading to very high $\rho_{\text{MIR}}$, and hence reaching $\rho_{\text{MIR}}$ is experimentally challenging.  We must remember that the MIR limit of $\sim 150 \; \mu\Omega \, \text{cm}$ has never been reached in any ordinary metal either (since metals would melt or vaporize long before reaching this high a resistivity), but it would be ludicrous to claim that copper is a bad metal because its measured resistivity never reaches the appropriate $\rho_{\text{MIR}}$.  Since cuprates are low-density metals, their $\rho_{\text{MIR}}$ values are simply too high to be observed experimentally in most situations.  This, by itself, should not be construed as evidence for a bad or strange metal -- it could simply be a reflection of $\rho_{\text{MIR}}$ being very high in most cuprates by virtue of the small Fermi surface. Having found evidence that the MIR limit is respected for at least one member of the cuprate family, we now raise the possibility that resistivity saturation might occur in other cuprates if sufficiently high resistivities of order $\sim 3-5 \; \text{m}\Omega \, \text{cm}$ can be accessed.  It is therefore possible that cuprates are not bad metals at all, but are just very highly resistive metals (still being below their MIR limit) because of strong quasiparticle scattering and a small Fermi surface.

We emphasize that our results and analysis do not in any way depend on any underlying assumption about the resistive scattering mechanism in our system.  We use the semi-classical Drude formula for estimating $\rho_{\text{MIR}}$ to compare with our measured resistivity, but this formula, our Eq. (\ref{mirbound}), is independent of any scattering mechanism.  Our key experimental finding is a saturation of the measured  resistivity around $\sim 1 \;\text{m}\Omega \, \text{cm}$ and the fact that this saturation value agrees with the estimated $\rho_{\text{MIR}}$ according to Eq. (\ref{mirbound}), with $k_F$ estimated from our Hall measurements of the effective carrier density, is simply an empirical secondary result.

In summary, we have demonstrated that the MIR limit is obeyed in as-grown samples of the electron-doped cuprate LCCO. These samples were chosen on account of their large room-temperature resistivity, enabling us to probe the MIR-regime at relatively low temperatures. The magnitude of the resistivity at which we observe saturation is substantially larger than the measured resistivity at 1000 K in hole-doped cuprates \cite{martinbsco, lscotakagi, bucherybco}. Further, the observed saturation consistently occurs near the estimated MIR limit, $\rho_{\text{MIR}}$, derived from a simple Drude model. Our results suggest the possibility that other cuprates may not be so-called ``bad metals'' at all, and may very well obey the MIR limit if probed at sufficiently high resistivity scales.

\begin{acknowledgments}
We thank P.R. Mandal and J.S. Higgins for fruitful conversations and experimental support. The experimental work is supported by the NSF under Grant No. DMR-1708334 and the Maryland Quantum Materials Center. S.D.S. is supported by the Laboratory for Physical Sciences.
\end{acknowledgments}

\bibliography{apssamp}

\end{document}